\documentclass[preprint]{aastex}
\usepackage{graphicx}

\usepackage{natbib}

\newcommand{\be}{\begin{equation}}
\newcommand{\ee}{\end{equation}}
\newcommand{\nn}{\mbox{} \nonumber \\ \mbox{} }
\newcommand{\ba}{\begin{eqnarray}}
\newcommand{\ea}{\end{eqnarray}}

\newcommand{\E}{{\bf E}}
\newcommand{\B}{{\bf B}}

\renewcommand{\v}{{\bf v}}

\renewcommand{\div}{{\rm \, \,div\,}}

\newcommand{\Bf}{{magnetic field}}
\newcommand{\Ef}{{electric field}}
\newcommand{\Bfs}{{magnetic fields}}
\newcommand{\NS}{neutron star}
\newcommand{\ms}{magnetosphere}
\newcommand{\mss}{magnetospheres}
\newcommand{\NSs}{{neutron stars}}
\newcommand{\EM}{{electromagnetic}}

\newcommand{\Lc} {{light cylinder}}
\newcommand{\BH} {{black hole}}

\newcommand\eg{{\it{e.g.}}}

\newcommand\lo{\mathrel{\raise.3ex\hbox{$<$}\mkern-14mu\lower0.6ex\hbox{$\sim$}}}
\newcommand\go{\mathrel{\raise.3ex\hbox{$>$}\mkern-14mu\lower0.6ex\hbox{$\sim$}}}

\begin{document}
\title{Electrodynamics  of  binary neutron star mergers
}

\author{Maxim Lyutikov}

\affil{
Department of Physics, Purdue University, 
 525 Northwestern Avenue,
 West Lafayette, IN
47907-2036 }

\begin{abstract}
We consider electromagnetic interaction and precursor emission of merging \NSs. Orbital motion of the magnetized neutron star may revive pair production within the common magnetosphere years before  the merger, igniting pulsar-like  magnetospheric dynamics. We identify  two basic scenarios: (i) only one star is magnetized (1M-DNS scenario)  and (ii) both stars are magnetized (2M-DNS scenario). Inductively created  {\Ef}s  can have component along the total \Bf\ (gaps)  and/or the \Ef\ may exceed the value of  the local \Bf. The key to the detection is orbital modulation of the emission.
If only one star is magnetized  (1M-DNS scenario)  the emission is likely to be produced along the direction of the \Bf\ at the location of the secondary; then, if the magnetic axis is misaligned with the orbital spin, this direction is modulated on the orbital period. For the 2M-DNS scenario,  the structure of the common \ms\ of the non-rotating  \NSs\  is complicated, with gaps, but no $E>B$ regions; there is strong orbital variations for the case of misaligned magnetic moments. For the same parameters of  \NSs\  the 2M-DNS scenario has intrinsically higher potential than the 1M-DNS one. The overall powers are not very high, $\leq 10^{45} $ erg s$^{-1}$;   the best chance to detect \EM\ precursors to the 
merging \NSs\ is if the interaction of their \mss\ leads to the production of pulsar-like  coherent radio emission modulated at the orbital period, with luminosity of up to $\sim 1$ Jankys at the time the merger.
\end{abstract}


\section{Introduction: expected precursors to DNS mergers}

The detection of gravitational waves associated with a GRB \citep{2017PhRvL.119p1101A} identifies merger of \NSs\ as  the central engine. It is highly desirable to detect any possible precursor to the main event.
At the time of the merger the \EM\ interaction of the \NSs\ can be considered as interaction of magnetized dipole(s), \S \ref{pre}.
\cite{2001MNRAS.322..695H} \citep[see also][]{2012ApJ...757L...3L} argued that magnetospheric interaction during double \NS\ (DNS) merger can lead to the production of \EM\ radiation. 
The  underlying mechanism  advocated in those works is a creation of inductive \Ef\ due to the relative  motion of
one NS, assumed to be unmagnetized, in the magnetic field of the companion \citep{GoldreichJulian,2011PhRvD..83f4001L}. 
Below we call it  the 1M-DNS case, \S \ref{1M-DNS}.

Typically simulations of the DNS merger with \Bf\ is done within the MHD approximation \citep[\eg][]{2013PhRvL.111f1105P,2013PhRvD..88d3011P,2014PhRvD..90d4007P,2015PhRvD..91h4038P,2017RPPh...80i6901B}. These simulations are well suited to understanding the overall structure of the interacting \mss. But they do no capture the dissipative/acceleration process.  In MHD (even with resistivity) the formation of gaps is prohibited. Thus, such simulations will miss the effects we discuss below -  formation of gaps and  dissipative regions with $E>B$. We develop  a complimentary \EM\ model of interacting \mss, based on the pulsar \ms\ model of \cite{GoldreichJulian}.

We expect that gaps forming in the \ms\ of interacting \NSs\   are not sustainable for a long time: if they create pairs the resulting charge separation screens the parallel \Ef, often in a non-stationary way \citep{2005ApJ...631..456L,2010MNRAS.408.2092T}. As  a result, the magnetospheric structure would generally evolve towards ideal MHD limit (without parallel component of the \Ef),  with a network of current sheets.  Simulations of \cite{2013PhRvL.111f1105P} do show strong magnetospheric interaction and formation of current sheets. Particles accelerated in reconnection events at these current sheets \citep[\eg][]{2018JPlPh..84b6301L} can also produce unstable distribution and coherent radio emission.

In this paper we  first consider the structure of the common \ms\ within the  1M-DNS model and then point out that the interaction of the \mss\ of the two {\it magnetized} \NSs\  -  2M-DNS scenario- will produce more powerful emission, \S \ref{2M-DNS}. Most importantly, in both cases we expect orbital-dependent emission pattern, that is a key to a possible future detection.

\section{Orbital resurrection of the rotationally dead}
\label{pre}
 
  A \NS\  after a time $t_{NS}$  since   birth will have a spin period \citep{GoldreichJulian}
 \be
 \Omega_{NS} \approx  \frac {c^{3/
       2} \sqrt {I_ {{NS}}}} {\sqrt {2} \sqrt {t_{NS}} B_ {{NS}} R_ 
{{NS}}^3} = 0.7 {\rm \, rad \,s}^{-1} t_{NS,9}^{-1/2},
\label{ONS}
 \ee 
 where the  surface \Bf\ $B_{NS}=10^{12}$ G is assumed, $t_{NS,9}$ is time in gigayears; $I_{NS} \approx 10^{45}$ g cm$^2$ is the moment of inertia, $R_{NS}=10$km  is \NS\ radius. Thus in a giga-year a star spins down to $\sim 10$ seconds. 
 
 The electric potential $\Phi_{spin}$ due to pulsar spin is \citep{GoldreichJulian}
 \be
 \Phi_{spin} \approx \frac{\Omega ^2 B_{{NS}} R_{{NS}}^3}{c^2} = 100\, t_{NS,9}^{-1} \,  GeV 
 \ee
  Typically, vacuum breakdown occurs at $\Phi \geq  100$ GeV \citep{1977ApJ...217..227F,1996ApJ...457..253D,Hibschman,2010MNRAS.408.2092T}. 
  Thus the merging \NS\ are likely to be dead pulsars, with no pair production within their \mss. \citep[There is an apparent exception to this, ][]{2018arXiv180900965T}

 At the orbital separation $r$ the time to merger $(-t)$ is \citep{MTW}
 \be
 -t\approx \frac{c^5 r^4}{ {( G M_{NS}) }^3}
 \ee
The light cylinder $R_{LC} =  c/ \Omega_{NS}$ becomes smaller that the orbital size at time $(-t_{LC})$ 
\be
(-t_{LC}) = 
 \frac{4 c^3 B_{{NS}}^4 R_{{NS}}^{12} t_{{NS}}^2}{G^3 {I_{NS}}^2
   M_{{NS}}^3} = 5 \times 10^{8} yrs
\ee
Thus for a few hundred million years the \NSs\ are within each other's \ms, so that their interaction can be considered as an interaction of  either a vacuum dipole with a highly conducting sphere (if only one star is magnetized) or two vacuum dipoles. At this point stars are separated by $5 \times 10^{10}\, t_{NS,9}^{1/2} {\rm cm}$.
(Also, the \Lc\ for the orbital motion when stars are separated by $r$  is located at
   \be
   R_{LC} \sim \frac{c r^{3/2}}{\sqrt{G M}}=
   \frac{G^{5/8} (-t)^{3/8} M_{{NS}}^{5/8}}{c^{7/8}}=
   2 \times 10^7 (-t)^{3/8}
   \ee
  This is much larger that the size of the orbit until the final merger.)

As the stars spiral in, the \mss\ can be revived due to relative motion of the magnet(s). 
On basic grounds \citep{2002luml.conf..381B,2006NJPh....8..119L}, if a system has typical \Bf\ B, internal velocity $v = \beta c$ and  typical size $R$,  the electric potential and \EM\ luminosity can be estimated as
\ba &&
\Phi \sim   \beta B R
\nn &&
L \sim \Phi ^2 c
\label{0}
\ea

In this paper we discuss two cases that both employ relations  (\ref{0}), but in somewhat different regime:  Single-magnetized DNS (1M-DNS) and Double-magnetized DNS (2M-DNS) mergers. The case of 1M-DNS is akin to a conductor moving in \Bf\ - there is then an induced \Ef, mostly close to the conductor's/{\NS}'s surface, that induces {\it a la} Goldreich-Julian surface charges, that create parallel \Ef, that leads to particle acceleration. In addition - this is different from the pulsar \ms\ case - regions with $E>B$ can be  created.  Like in case of pulsars, regions with  $\E\cdot \B \neq 0$ and $E>B$ will lead to particle acceleration, pair production, and shortening of the acceleration \Ef, most likely in a non-stationary way \citep{2005ApJ...631..456L}.

The case of 2M-DNS involves interacting relativistic  \mss\ - here the stars play a role of the point dipoles that produce \EM\ fields in their surrounding. Those \EM\ fields can form gaps  and/or reconnecting current sheets, leading to energization,  acceleration of particles, and production of non-thermal  and possibly coherent radio emission.

The power and the available potential in both cases  of 1M-DNS and 2M-DNS  scenarios can be estimated using (\ref{0}) with typical \Bf\ due to a dipole at the distance of orbital separation $B \sim B_{NS} (r/R_{NS})^{-3}$ and typical radius $R=R_{NS}$ in case 1M-DNS (Eq. (\ref{L1})) and $R=r$ in the case 2M-DNS (Eq. (\ref{L2}).

 Typically, there is enough potential (see Eqns.  (\ref{L1}) and (\ref{L2}) ) to start pair production at   times
   before merger
   $(-t) \sim 10^{12} $ seconds, $\sim 5 \times 10^4$ years.  At that moment the stars are separated by $\sim 5 \times 10^9$ cm. Starting this moment we may expect some \EM\ signal from the system.

As we discuss below, most of the  common \ms\ remains open, so that we do not expect large stationary  regions filled with plasma, like in the case of pulsar's closed field lines \citep{GoldreichJulian}. Also, pair production is likely to be intermittent. Thus, a vacuum approximation is expected to be a  reasonable approximation.

Let us consider the \EM\ structure of  two interacting vacuum  DNS \mss. Following  of the classic work of \cite{GoldreichJulian},  we will find that orbiting \NSs\ create electric potentials along the combined {\Ef}s. This serves as a first-order approximation to the expected  nearly ideal plasma, that  at the same time has special accelerations regions  with $\E\cdot \B  \neq 0$.

Let us next discuss cases 1M-DNS and 2M-DNS in detail.

\section{Interacting DNS \mss: 1M-DNS case}
\label{1M-DNS}

If only one star is magnetized, we are dealing with a metal sphere (second \NS) moving through the \Bf\ of the primary. This was the case originally considered by \cite{2001MNRAS.322..695H}. Below in this section we provide a more detailed description of the resulting interaction.

\subsection{1M-DNS: estimates of power}
If a \NS\ is moving in the field of a primaries' dipolar \Bf\ at orbital separation  $r$, 
the induced potential and the corresponding powers are \citep{2001MNRAS.322..695H,2011PhRvD..83l4035L}
\ba &&
\Phi_1 \sim B R_{NS}= \frac{B_{{NS}} R_{{NS}}^4 \sqrt{G M_{{NS}}}}{c r^{7/2}} =
 10^{18} (-t)^{-7/8} \, {\rm \, eV}
   \nn &&
L_1 \sim  \frac{G B_{{NS}}^2 M_{{NS}} R_{{NS}}^8}{c r^7}=3 \times { 10^{41} }{(-t)^{-7/4}}\, {\rm \, erg \, s^{-1}}
\label{L1}
\ea
where in the last relations the time to merger $t$ is measured in seconds.
(Index $1$ indicates here that the interaction is between single magnetized \NS\ and unmagnetized one.)

\subsection{1M-DNS - internal structure of the \ms}

Consider a highly conducting unmagnetized \NS\ moving through the \Bf\ of the companion. The expected parallel \Ef\ that would develop in the system will be largest near the surface of the companion (unmagnetized) \NS. Hence, for  simplicity we approximate the dipolar \Bf\ of the primary at the location of the secondary as a straight vertical field. This is a good approximation if the sizes of the stars are much smaller than the orbit. 

\begin{figure}[h!]
\includegraphics[width=.99\columnwidth]{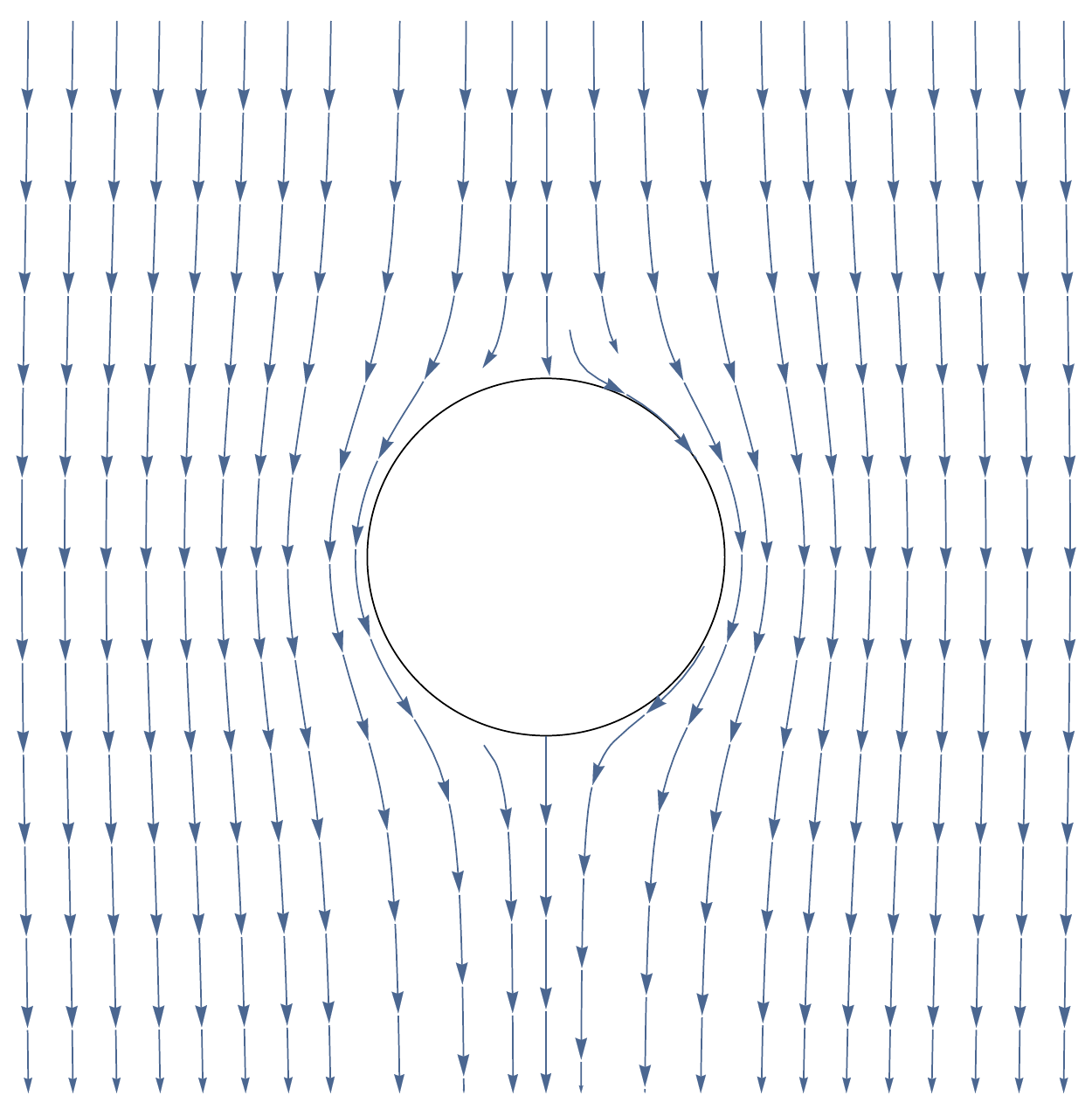} 
\caption{Expulsion of the \Bf\ by a conducting sphere.}
 \label{Bexpulsa}
\end{figure}

The \NS\ is highly conducting, and can be approximated as a conducting metal sphere, so that the \Bf\ does not penetrate, Fig. (\ref{Bexpulsa}).
The induced field is
\ba &&
B_r =-\frac{R^3}{r^3} \cos \theta B_0
\nn &&
B_\theta= 
 \frac{R^3}{r^3} \sin \theta  B_0
   \label{Bind}
   \ea
while the total
\Bf\ is 
\ba &&
B_r =-  \left(1-\frac{R^3}{r^3}\right) \cos \theta B_0
\nn &&
B_\theta= 
\frac{1}{2} 
   \left(2+ \frac{R^3}{r^3}\right) \sin \theta  B_0
   \label{BindTot}
   \ea
   
   The conducting sphere is sliding through the \Bf\ with the relative velocity
   \be
   \v = 2 \Omega \times {\bf r}
   \ee
   (here $r$ is a semi-major axis).
As a result \Bf\ in the observer frame is time-dependent: there is non-zero $\partial_t \B$, which would produce \Ef. The resulting \Ef\ will generally have a component parallel to the \Bf. 

Below, in \S \ref{orth} we first consider a case of the secondary \NS\ moving orthogonally (in the equatorial plane of the primary)   through the \Bf\ of the primary. Later, \S \ref{oblique}, we generalize to oblique propagation.

\subsection{1M-DNS - motion of the secondary  in the  magnetic equatorial plane of the primary}
\label{orth}
let us first consider the   case when the secondary unmagnetized  star is moving perpendicular to the \Bf\ lines. In Eq. (\ref{BindTot}) make a transformation to the observer frame by writing $y \rightarrow y+\beta_0  t$. Time derivative of (\ref{BindTot}) (evaluated at arbitrary $t=0$) is
\be
\partial_t \B =
\left\{-\frac{9}{4} \sin 2 \Theta \sin \phi ,\frac{3}{2} \cos 2 \Theta \sin (\phi
   ),\frac{3}{2} \cos \Theta \cos \phi \right\}
    \frac{ \beta_0 B_0 R^3}{ r^4}
   \ee
   (in $\{r,\theta,\phi\}$ coordinates.
This equals minus the curl of the induced \Ef. We find a vacuum \Ef\ with $\div \E=0$
\be
\E_{ind}=-\left\{\frac{1}{2} \sin \Theta \cos \phi ,-\cos \Theta \cos \phi ,\frac{1}{4}
   (3 \cos (2 \theta )+1) \sin \phi \right\}
\frac{ \beta_0 B_0 R^3}{ r^3}
\label{Eind}
\ee
(alternatively, we could  do a Lorentz boost of the induced \Bf\ (\ref{Bind}), giving the same result). 

The induced field $\E_{ind}$ (\ref{Eind}) has a component normal to the surface of the NS, that will produce  surface charge density  $\sigma$
\be
\left.E_{ind,r}\right|_{r=R} =-\frac{1}{2} \sin \Theta \cos \phi  B_0 v =  4 \pi \sigma
\ee
The surface charge will produce an \Ef\
\be
\E_{s}=\left\{ \frac{1}{2} \sin \Theta \cos \phi ,-\frac{1}{4} \cos \Theta \cos \phi ,\frac{1}{4}
   \sin \phi \right\}
\frac{ \beta_0 B_0 R^3}{ r^3}
\ee
The total field outside is then
\be
\E_{tot} =\E_{ind}-\E_{s}= \left\{-\sin\Theta \cos \phi  ,\frac{5}{4} \cos \Theta \cos \phi ,\frac{1}{4} (2+3 \cos 2 \Theta) \sin
   \phi \right\}
   \frac{ \beta_0 B_0 R^3}{ r^3}
   \label{etot}
\ee

The total magnetic and the electric  fields (Eqns. (\ref{BindTot}) and (\ref{etot})) have two important properties. They have non-zero first \EM\ invariant  $\E \cdot \B$ and the second invariant $B^2-E^2$ can change sign. 

\begin{figure}[h!]
\includegraphics[width=.99\columnwidth]{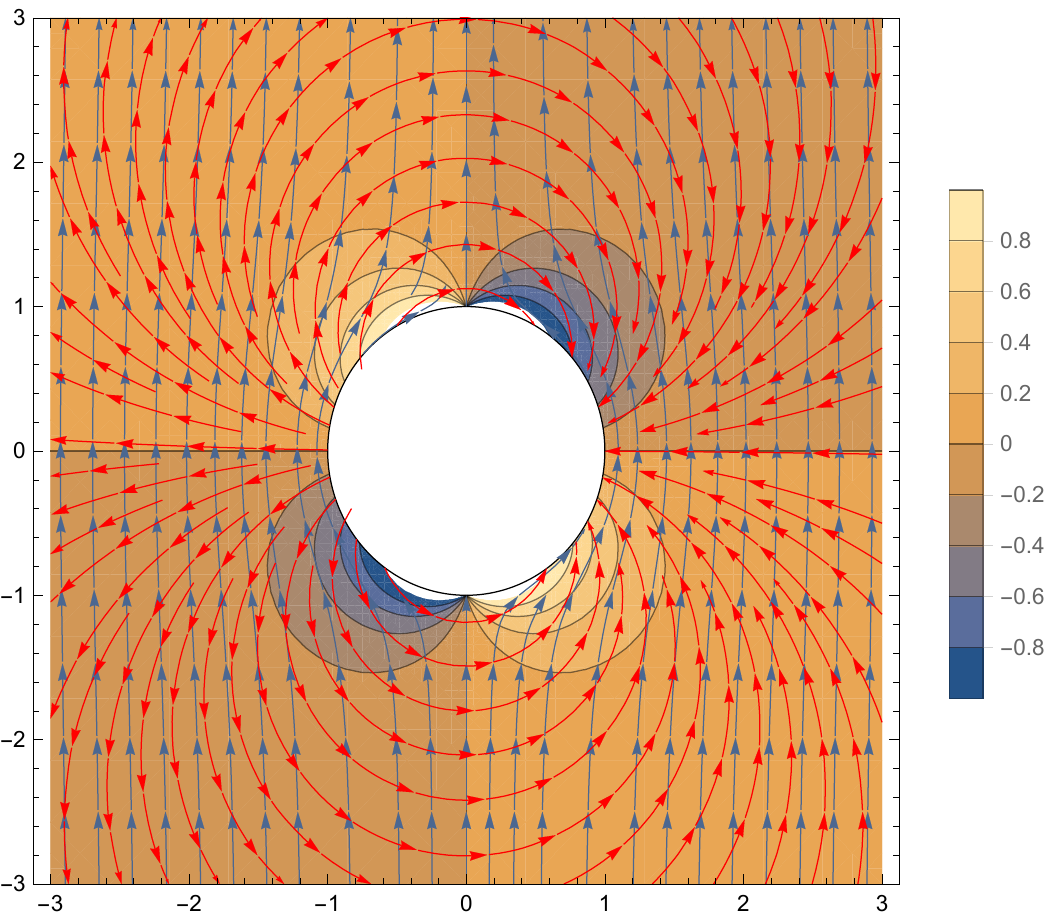} 
\caption{Magnetic field lines (blue arrows), total \Ef\ (red arrows) and the value of $\E_\parallel$ (in color)  in the plane $y=0$ for a metal sphere of radius $R$  moving in constant \Bf.  Electric field is measured in units  $B_{NS} \beta R^3$ and coordinates in $r/R$.}
 \label{Eparao}
\end{figure}

The parallel component of the \Ef\
 equals
\be
E_\parallel=-  \frac{3}{2 \sqrt{2}}\frac{\sin \Theta \cos \Theta \left(6- \frac{R^3}{r^3}\right) \cos (\phi
   )}{\sqrt{8 \left( 1- \frac{R^3}{r^3}\right)^2+ 6 \left( 4- \frac{R^3}{r^3}\right) \frac{R^3}{r^3} \sin^2 \Theta}}
B_0 \beta \left( \frac{R}{r}\right)^3.
\ee
 Fig. \ref{Eparao} shows that there are regions of non-zero parallel \Ef, peaking in value close to the \NS\ surface. If plasma is generated in these regions, the outflows will be directed mostly along the external \Bf, passing close to the poles.

The parallel \Ef\ is maximal in the $y=0$ plane ($\phi=0,\pi$). At each $r\neq R$ the maximal values of $E_\parallel$ is reached at $\cos^2 \theta = ( 2 r^3+R^3)/(4 r^3-R^3)$. At
 $r\rightarrow R$ this corresponds to $\theta= 0,\pi$, where the total \Bf\ is zero.
 
 The invariant $B^2-E^2$
 equals
 \ba &&
 B^2-E^2 = 
 \frac{1}{16}
 \left(
 \left(16 \left(1-\frac{r^3}{R^3}\right)^2-25 \beta ^2 \cos
   ^2(\phi )\right)\cos ^2(\Theta ) +
   4  \left(\left(\frac{2 r^3}{R^3}+1\right)^2-4 \beta
   ^2 \cos ^2(\phi )\right)\sin ^2(\Theta )
   \right.
   \nn &&
   \left.
   -\beta ^2 (3 \cos (2 \Theta )+2)^2 \sin ^2(\phi )
   \right)
 \frac{R^6}{r^6}    B_0^2
   \ea
   
\begin{figure}[h!]
\includegraphics[width=.4\columnwidth]{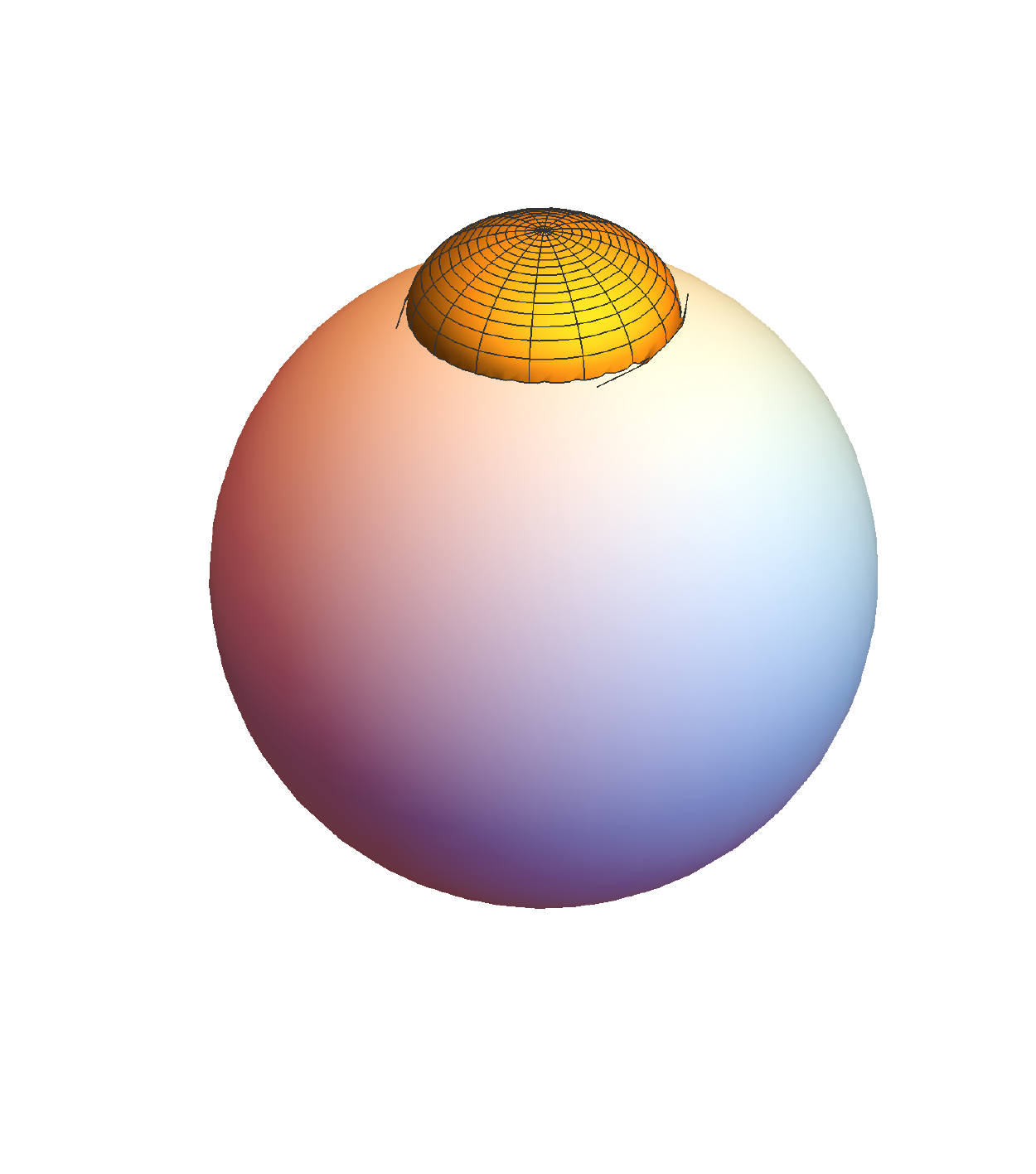} 
\includegraphics[width=.49\columnwidth]{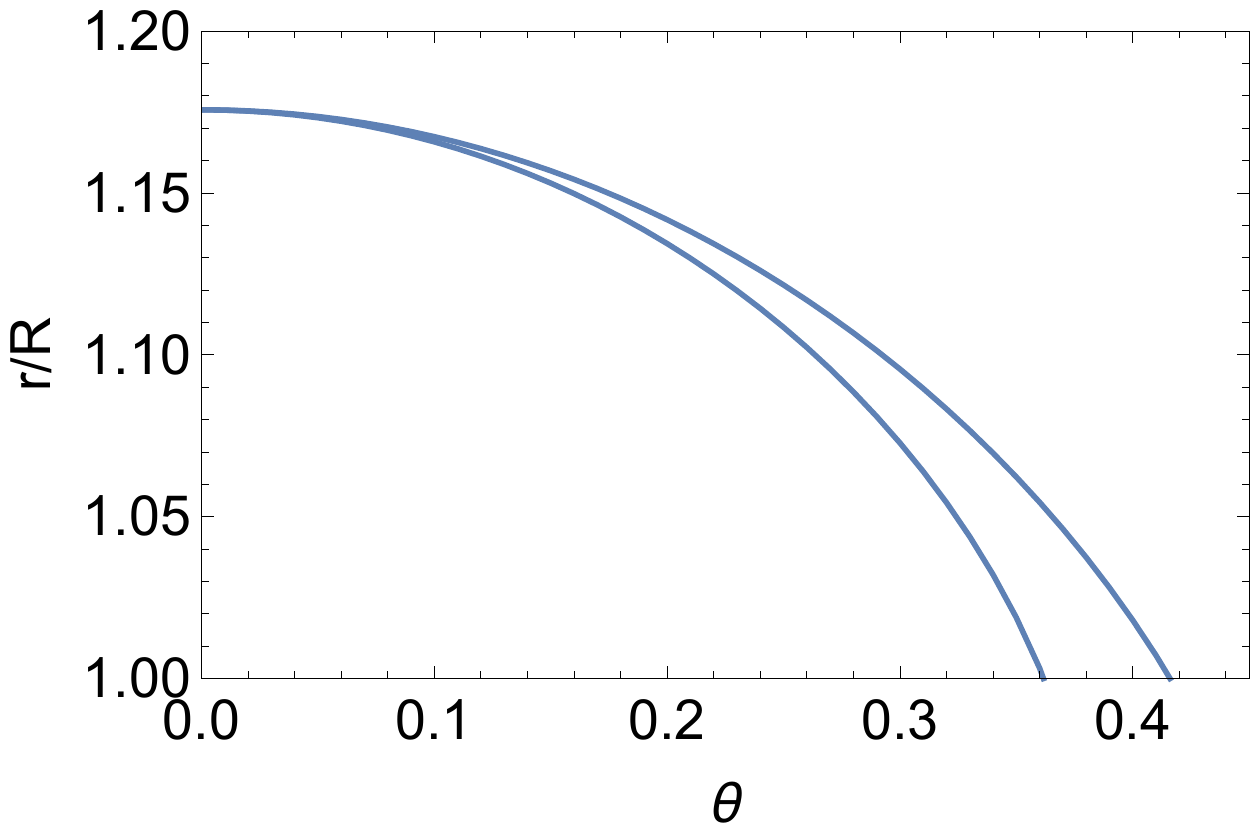} 
\caption{
(Left panel) 3D rendering of the regions with $B<E$. The region is slightly non-circular:  the right panel shows cross-section in the $\phi=0$ plane (top curve) and    $\phi=\pi/2$ plane  for $\beta =0.5$.}
 \label{anlge2ndinv}
\end{figure}

   It becomes negative within a dome-like structure near the poles. On the  surface of the \NS\ it is $<0$ within a region 
   \be
   \cos ^2\phi= \frac{36 \sin ^2(\Theta )-\beta ^2 (3 \cos (2 \Theta )+2)^2}{3 \beta ^2  \sin ^2(\Theta ) (6 \cos (2
   \Theta )+11)}    \label{thetaC}
   \ee
   (for each  value of $\phi$ Eq.  (\ref{thetaC}) gives the  value of $\theta$ so that for smaller $\theta$ (closer to the pole) we   have $B<E$, see Fig. \ref{anlge2ndinv}).
   Note that even in the $\phi=\pi/2$ plane, where  the \Ef\ is zero, the \Bf\ is also zero, so that the the second invariant remains negative close to the poles. At
 the poles,  $\theta= 0, \pi$,  on the surface  $ B^2-E^2= - (25/16) B_0^2 \beta^2$; the region extends to $r/R= (1+(5/4) \beta)^{1/3}$.

It is expected that both the conditions $\E\cdot\B \neq 0$ and $B^2<E^2$ will lead to particle acceleration. This might create conditions favorable for the generation of coherent radio emission. Without a particular global model of particle acceleration we cannot predict in detail the expected radio signal. Both the  region $B<E$  and the region with the largest $E_\parallel$  are  concentrated near the pole, so that in the case of motion  in the equatorial plane, no variation is expected. 

The above results can be compared with the case of a \BH\ moving through straight \Bf\ \citep{2010Sci...329..927P,2011PhRvD..83f4001L}. In that case two dual jets are generated. As Fig. \ref{Eparao} demonstrates, in the case of moving metal sphere the configuration looks similar: there are four regions with high parallel \Ef. 

\subsection{1M-DNS - oblique motion of the secondary}
\label{oblique}

Above, in \S \ref{orth} we considered the motion of the secondary unmagnetized \NS\ in constant \Bf, a case applicable to when the magnetic moment of the primary is along the orbital spin. Next we generalize it to the oblique orbits of the secondary (inclined magnetic moment of the primary).

In the case of the oblique motion (not orthogonal to the spin) we can separate the motion  along and across the \Bf.
For the part of motion along the external \Bf, we find
\be
E_\phi= -\frac{3  \beta B_0 R^3 \sin (2 \theta )}{4 r^3}
\ee
Thus, this does not create a component  of \Ef\ along the \Bf. Also, the second \EM\ invariant
\be 
B^2-E^2 = \left( \frac{12 \cos (2 \theta )+20}{32 r^6}+\frac{9 v^2 (\cos (4 \theta )-1)}{32 r^6} \right) B_0^2 R^6
\ee
 is always positive.

Thus what counts for the producing emission regions ($\E\cdot \B \neq 0$ or $B<E$) is the component of the velocity perpendicular to the \Bf. Since the primary is assumed non-rotating, only the component of the velocity perpendicular to the magnetic moment of the primary is of importance. As we argued above, the emission is likely to be produced along the \Bf\ at the location of the secondary.

Consider a dipole of the primary inclined by the angle $\Theta$ with respect to orbit normal, located at the origin, $\{0,0,0\}$. The secondary at time $t$ is located at 
$ x_0 \{\cos \Omega t ,\sin \Omega t,0\}$ (in Cartesian coordinates $\{x,y,z\}$, where $\Omega$ is the orbital frequency. At the location of the secondary, the direction of the primary's \Bf\ is
\be
\B_{1-2}  =
\left\{-{\sin \Theta (3 \cos2 \Omega t +1)},
   -{3 \sin \Theta \sin2 \Omega t },
{2 \cos \Theta}\right\} \frac{1}{{\sqrt{-6 \cos 2 \Theta \cos ^2\Omega t
   +3 \cos2 \Omega t +7}}}
   \ee 
in $\{x,y,z\}$ coordinates. The direction of emission varies on the orbital time scale.  
Thus,  if emission is generated near the secondary along the local direction of the magnetic field of the primary we expect modulation of the direction on orbital time scale.

\section{Interacting DNS \mss: 2M-DNS case}
\label{2M-DNS}

\subsection{2M-DNS -  general notes}

Next we study the  interaction of two dipoles that are moving with respect to each other with the (Keplerian) velocity $v = \beta c$. We can identify 4 basic geometries:
 (i-a) magnetic moments aligned with the orbital spin axis;  (i-b) magnetic moments anti-aligned with the orbital spin axis; 
 (ii-a) one magnetic moment is aligned with the orbital spin axis and another is along the line connecting the two {\NS}s;
 (ii-b) one magnetic moment is aligned with the orbital spin axis and another is orthogonal both to the orbital axis and the line connecting the two {\NS}s.
 Surely, there more complicated structures when both  magnetic moments are misaligned, but these four, we think, cover the basics. 
 
 Qualitatively, cases i-a and i-b  produce nearly  time independent signals  and are less of interest (in this case the procedure  described below implies that the maximal emission is produced along the $z$-axis, though at different orbital phases the line of sight does pass through different regions). Cases ii-a and ii-b are dynamically related  - a system of  {\NS}s
 where one \NS\ is aligned with the orbital spin and the spin of the order is in the orbital plane will periodically evolve from ii-a to ii-b. As we demonstrate below, this will lead to considerably different magnetospheric structures and possibly time-dependent radiative signatures.
 
 Consider two {\NS}s of same masses and producing in the surrounding space   \Bfs\ $\B_1$ and $\B_2$. 
 Assume that the  {\NS}s are non-rotating - typical merger times are much longer than the typical spin-down times and orbital synchronization is not important \citep{1992ApJ...400..175B}. In the center of mass the star 1 is moving with velocity ${\bf v}_1= \Omega \times {\bf r}_1$ and star 2 is moving with velocity ${\bf v}_2= \Omega \times {\bf r}_2=- {\bf v}_1 $.  Neglecting relativistic effects,  they create the total \Bf\ $\B= \B_1+\B_2$ and the \Ef\ $\E= - {\bf v}_1 \times \B_1 - {\bf v}_2 \times \B_2=  - {\bf v}_1 \times ( \B_1- \B_2)$.  This total \Ef\ has a component parallel to the total \Bf:
 \be
  \E\cdot \B= -( {\bf v}_1 \times ( \B_1- \B_2)) \cdot ( \B_1+\B_2)=
-  ( {\bf v}_1 \times  \B_1) \cdot \B_2 + ( {\bf v}_1 \times  \B_2) \cdot \B_1
= -2 ( {\bf v}_1 \times  \B_1) \cdot \B_2
,
\label{EdotB}
  \ee
  which is generally non-zero.
    Thus, regions of non-zero parallel \Ef\ will be created within the interacting \mss.

\subsection{2M-DNS - estimates of power}
In this paper we point out that {\it magnetospheric interaction} of two magnetized \NSs\ can generate larger luminosity that the case of  one star moving in the field of the other considered by  \cite{2001MNRAS.322..695H}, see also  \S \protect\ref{1M-DNS}.
  Qualitatively, the relative motion of two \NSs\ creates large scale \Ef\ that  generally has a component parallel to the total \Bf. These large scale {\Ef}s are on the orbital scale, so that
\ba  &&
\Phi_2 \sim   \beta B r=\frac{  B_{{NS}} \sqrt{G M_{{NS}}} R_{{NS}}^3}{c
   r^{5/2}} = 
\frac{c^{17/8}  B_{{NS}} R_{{NS}}^3}{ (-t)^{5/8}
   ( G M_{NS})^{11/8}}=
   4 \times 10^{18} (-t)^{-5/8} 
  {\rm \, stat-V}
\nn  && 
L_2 \sim \frac{ B_{{NS}}^2 G  M_{{NS}} R_{{NS}}^6}{c  r^5}    =
\frac{c^{21/4} B_{{NS}}^2 R_{{NS}}^6}{
   (-t)^{5/4} ( G M_{NS})^{11/4}}=
 6 \times { 10^{42} }{(-t)^{-5/4}}\, {\rm \, erg \, s^{-1}}
    \label{L2}
\ea
(Index $2$ indicates here that the interaction is between two magnetized \NS.)
The ratio of luminosities of the models  1M-DNS  and 2M-DNS is 
\be
\frac{L_2}{L_1}  = \left( \frac{G M}{c^2 R_{NS}}\right)^{3/2} \sqrt{\frac{(- t) c}{R_{NS}}} \approx 16 \sqrt{-t}
\ee
Thus $L_2$ dominates $L_1$ prior to merger.

  The power (\ref{L2}) is fairly small until the  last few minutes.  Even at the time of a merger, with $t\sim 10^{-2}$   seconds the corresponding power is only
  $L \sim  10^{45}{\rm \, erg \, s^{-1}}$ - amplification of \Bf\ is needed to produce a typical   GRBs with  $L \sim  10^{50}{\rm \, erg \, s^{-1}}$.

If  at the precursor stage  the majority of the power comes out at soft photons with $\epsilon_{ph} \sim 1 $ keV, the expected high energy photon flux  at the Earth is
\be
F=  \frac{L_2}{4 \pi d^2 \epsilon_{ph} } \sim   3 \times 10^{-3} {(-t)^{-5/4}} d_{100 \, {\rm \, Mpc}}^{-2} \,  {\rm \, phot. s^{-1} cm^{-2}}
\ee
It is smaller by few orders of magnitude than the detection limits of   high energy satellites like Swift \citep{Swift}.

The best case, we believe, is if a fraction of the power (\ref{L2}) is put into radio.  If a fraction of $\eta_R$ of the power is put into radio, the expected signal then is 
\be
F_R \sim \eta_R  \frac{L_2}{4\pi d^2  \nu}\approx 0.1 {\rm \, Jy}\,  \eta_{R,-5} (-t)^{-5/4}
\label{FR}
\ee
This is a fairly strong signal that could be detected by  modern radio telescopes. Of course, it's a transient source - but the power (\ref{FR}) is fairly large for a long time before the merger.

  \subsection{Case i: magnetic moments parallel to the orbital spin}
  
\begin{figure}[h!]
\includegraphics[width=.49\columnwidth]{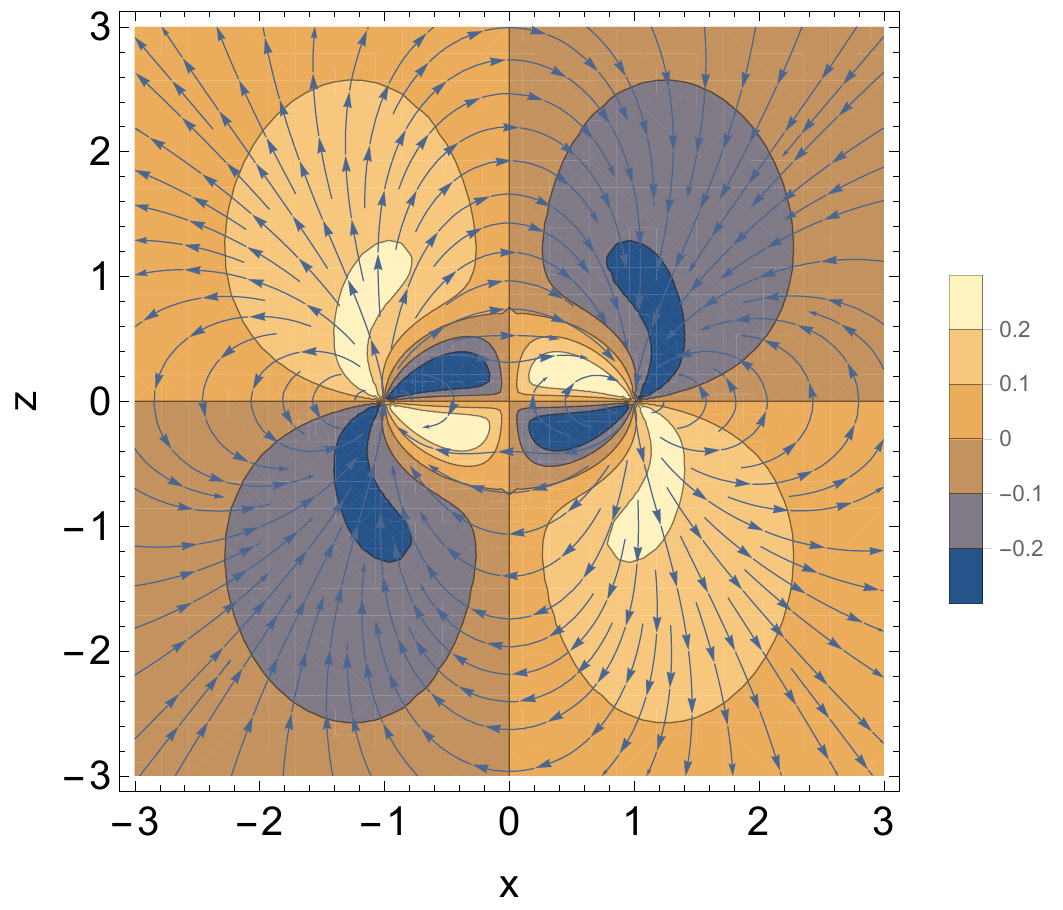} 
\includegraphics[width=.49\columnwidth]{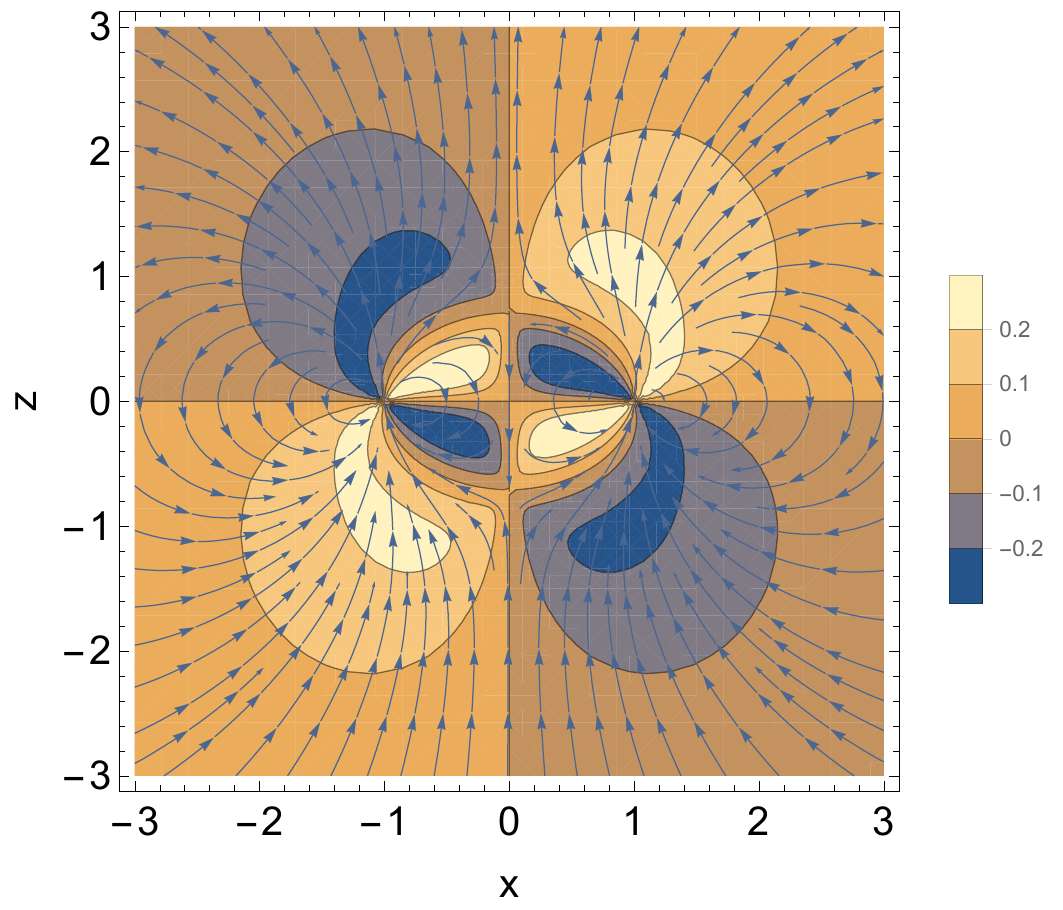} 
\caption{Case i: magnetic field lines (arrows)  and the value of $\E_\parallel$ (in color) for aligned (left) and misaligned (right) cases in the plane $y=0$.  Electric field is measured in units  $B_{NS} \sqrt{G M_{NS}} R_{NS}^3 /( c r^{7/2})$ where $r$ is the semi-major axis of the orbit.}
 \label{i}
\end{figure}

  Let us assume that the stars are located at coordinate $ x= \pm x_0$ and are orbiting each other in the $x-y$ plane (so the orbital spin is along z-axis). (We neglect the sizes of the stars - the corresponding \EM\ effects  that take into account the size of the \NSs\ were considered in \S \ref{1M-DNS}). Below we will mostly plot 2D images in the $\{x-z\},\, y=0$ plane (defined by the spin axis and the line connecting the stars, in the plane of the stars) and the  $\{y-z\},\, x=0$ plane 
  (orthogonal to the  line connecting the stars, passing through the center of mass). Results for  the parallel case i  are plotted in  Fig. \ref{i}.
  
 In the case of  orbiting \NSs\ with magnetic moments aligned  with the orbital spin, there is some time dependence in the structure of the \ms, but our prescription for estimating the preferred  direction of the emission (see below) consistently predict emission along the spin axis, and hence no time variation. Nearly constant  weak high energy or radio emission will be hard to detect observationally.
  
 \subsection{Case ii: one magnetic moment parallel to the orbital spin and  another in the plane of the orbit}

\begin{figure}[h!]
\includegraphics[width=.49\columnwidth]{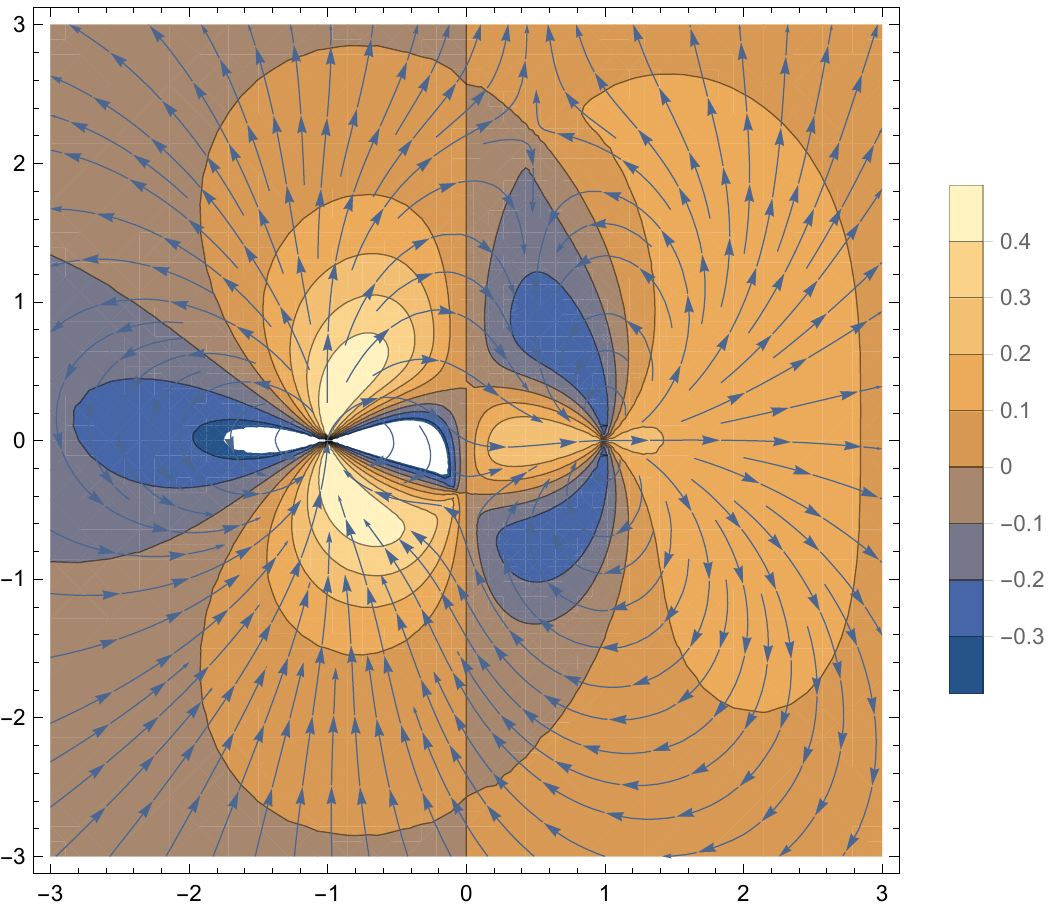} 
\includegraphics[width=.49\columnwidth]{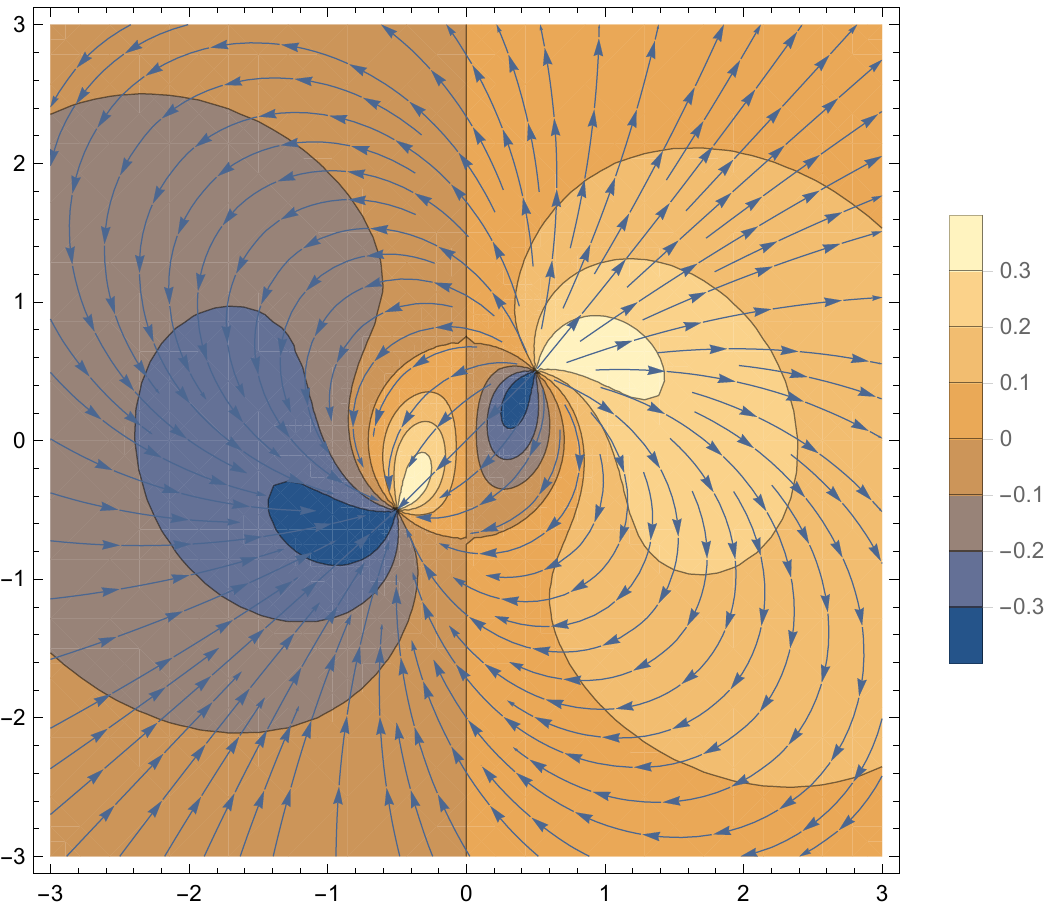} 
\caption{Case ii: magnetic field lines (arrows)  and value of $\E_\parallel$ (in color) for case ii-a  in the plane $y=0$ (containing the orbital spin and the two {\NS}s). (left) and case ii-b  (right) in the plane $x=0$. }
 \label{ii}
\end{figure}

For this configuration the magnetic and {\Ef}s are plotted in Fig. \ref{ii}.

   \subsection{Case ii: time dependence}
   
 In contrast to the cases-i, the  cases-ii are  highly time dependent, as the system will evolve from ii-a, to ii-b, to ``minus'' ii-a,  ``minus'' ii-b (by ``minus''  we mean the relative orientation of the magnetic moment with respect to the line of sight).
 Thus, if the spins of the {\NS}s are misaligned with the orbital spin, the structure if the \ms\ will change quasi-periodically on the orbital time scale. This may result in periodically changing emission. 
 
\begin{figure}[h!]
\includegraphics[width=.45\columnwidth]{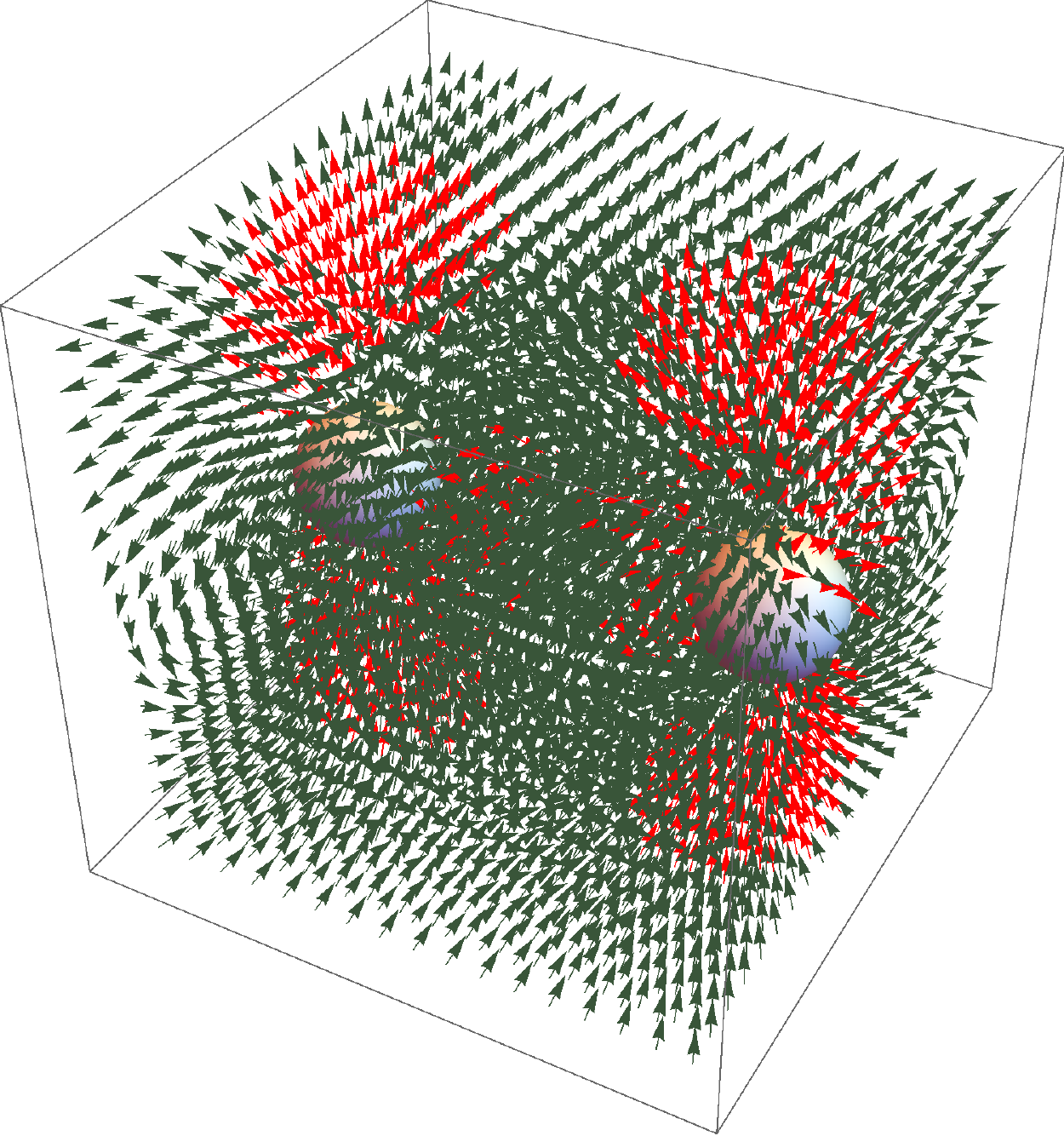} 
\includegraphics[width=.45\columnwidth]{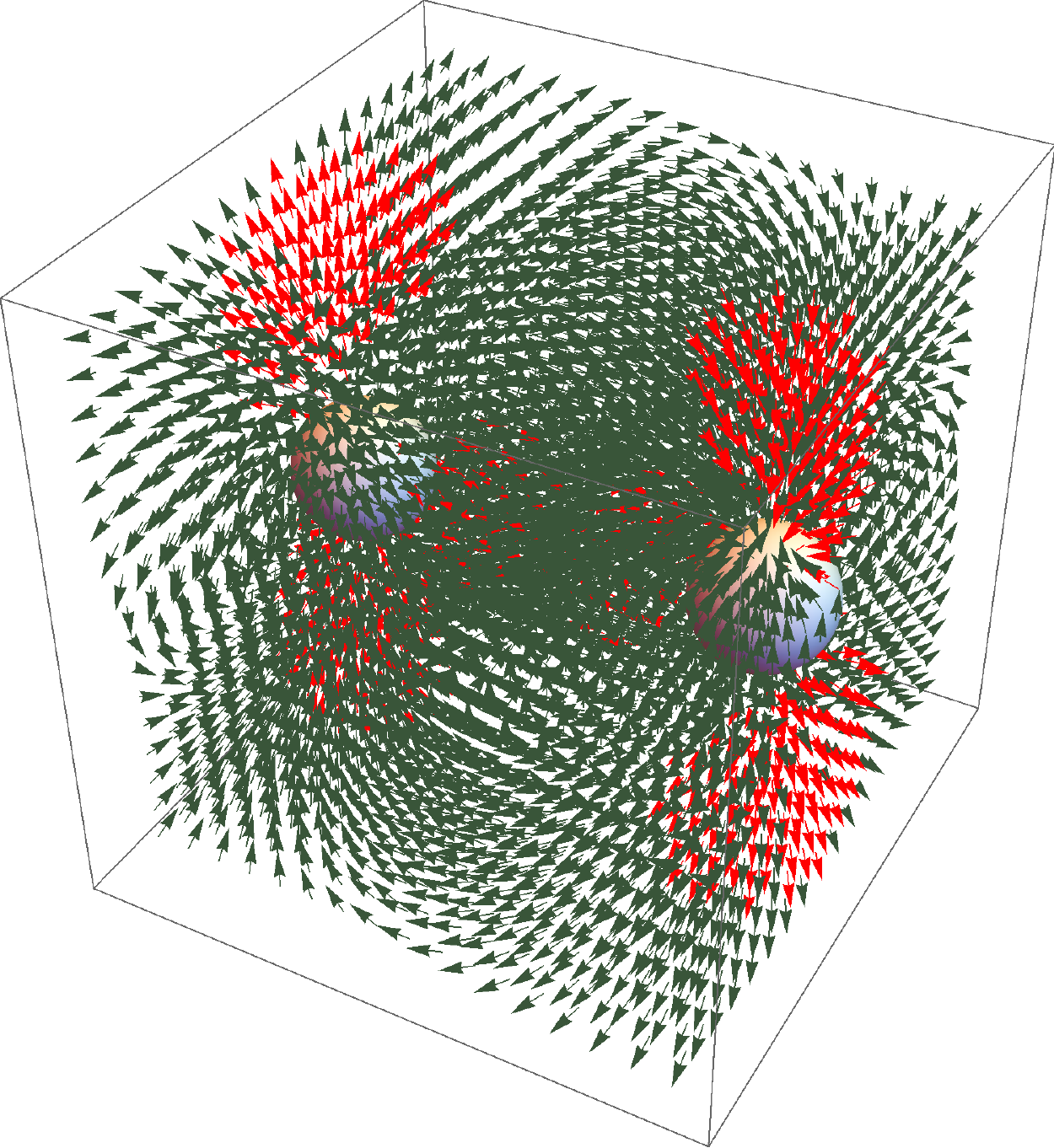} \\
\includegraphics[width=.45\columnwidth]{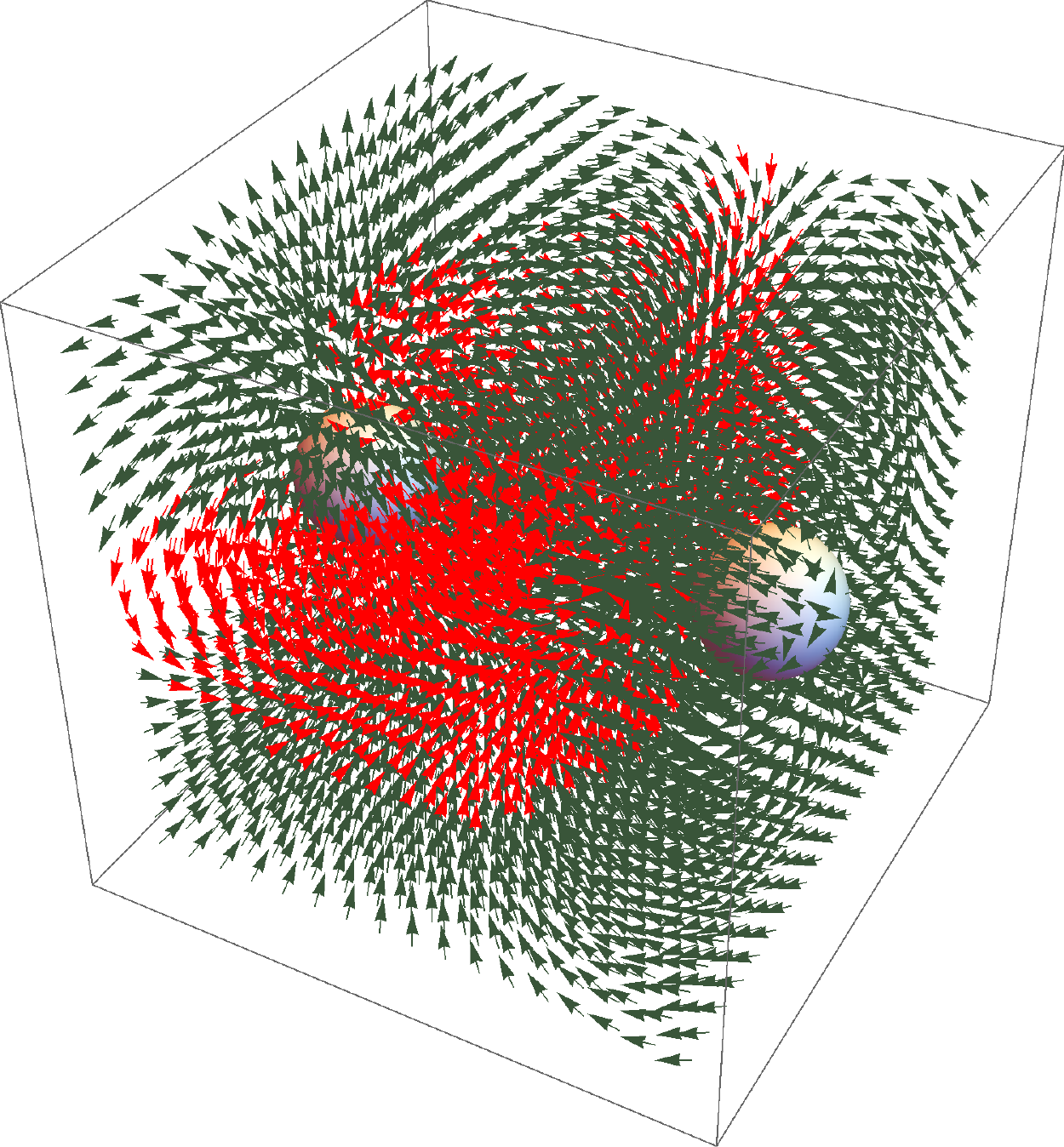} 
\includegraphics[width=.45\columnwidth]{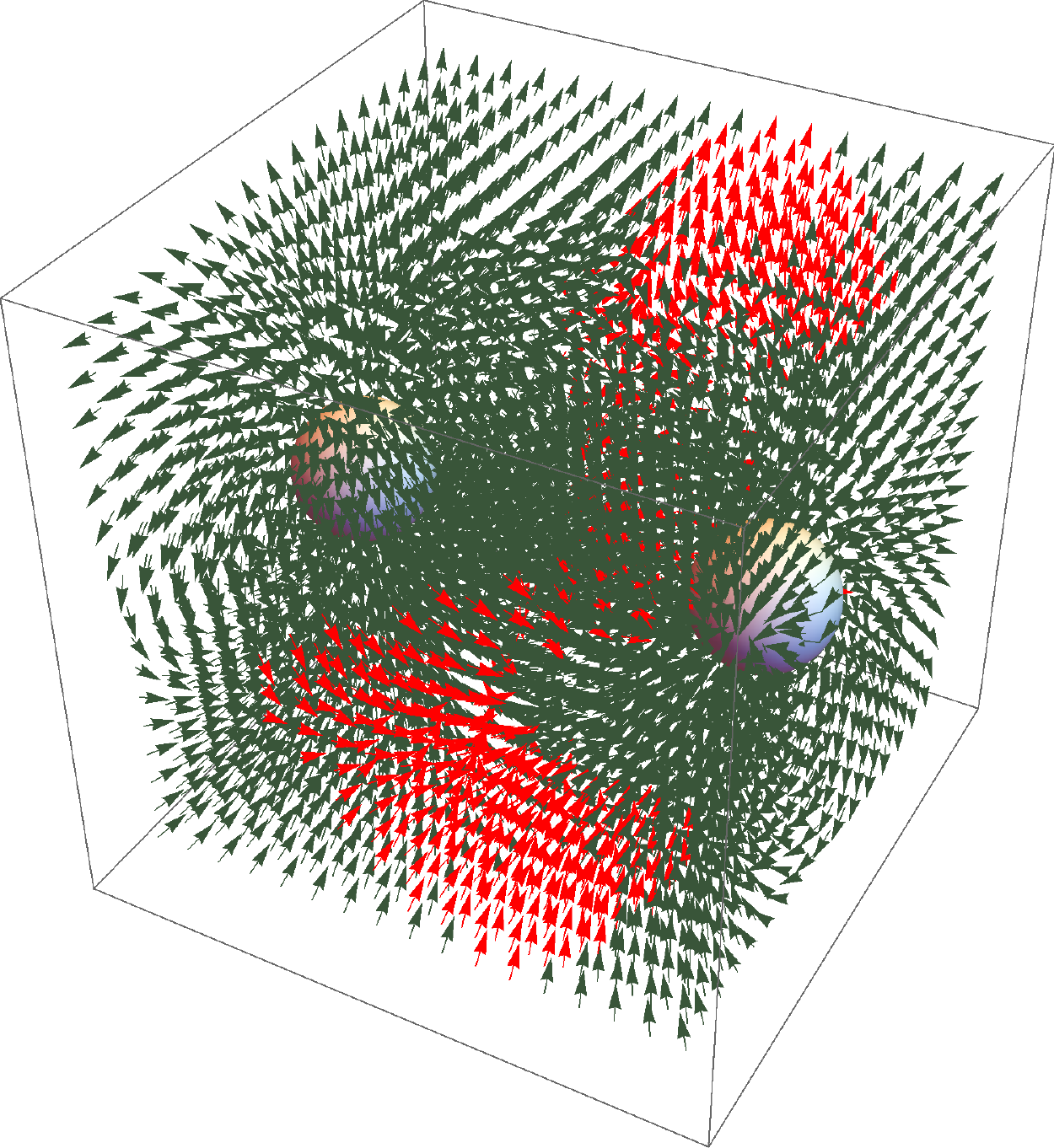} 
\caption{3D rendering of the common \ms\ of the interacting \NSs\   for four basic orientation of dipoles. Shown are aligned (top left), anti-aligned (top right),  and two orthogonal cases: one of the dipoles in the plane of the orbital spin and line connecting the stars ($x-z$ plane; low left), and  perpendicular one (low right). Highlighted in red are regions with high parallel \Ef. (Since in the orthogonal cases the parallel \Ef\ is larger in absolute value than in the parallel/anti-parallel cases different criteria were used  to highlight large $E_\parallel$ regions in different configurations.}
 \label{Magnetosph}
\end{figure}

\begin{figure}[h!]
\includegraphics[width=.49\columnwidth]{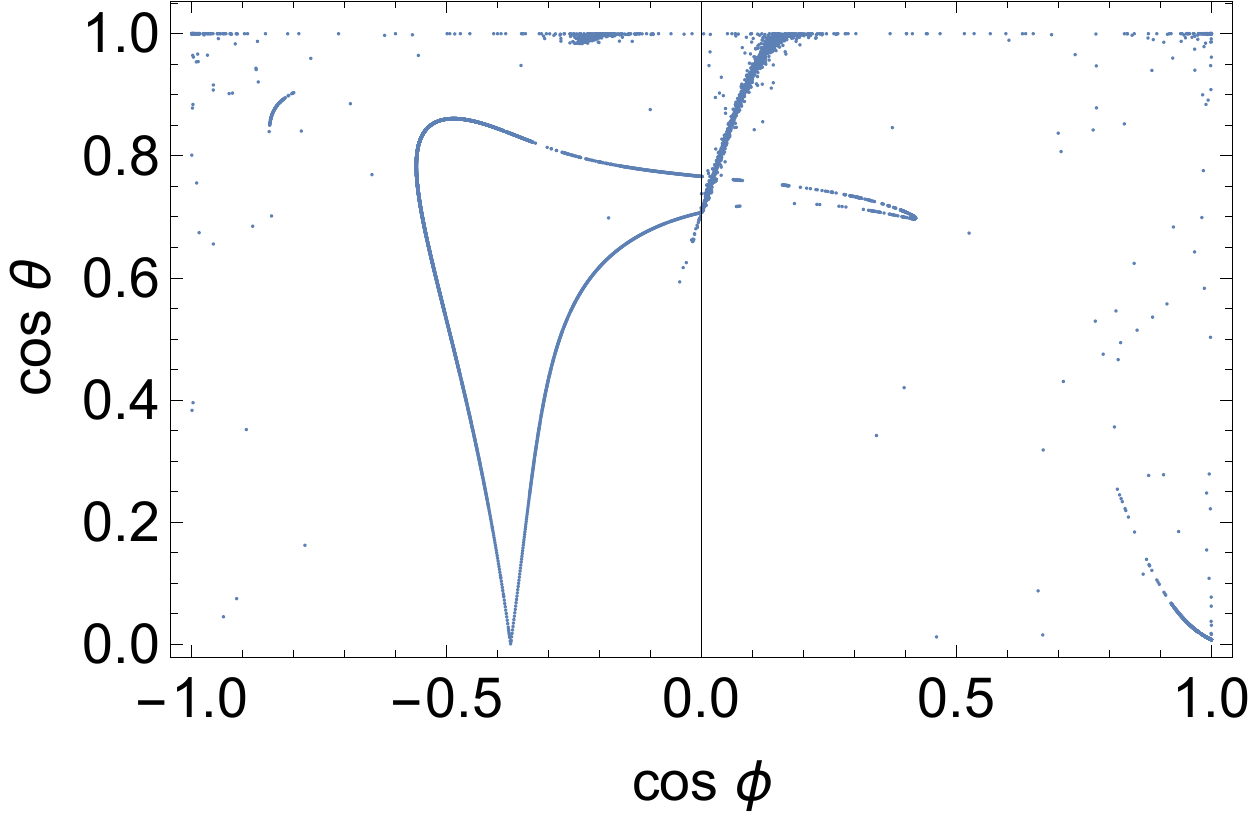} 
\includegraphics[width=.49\columnwidth]{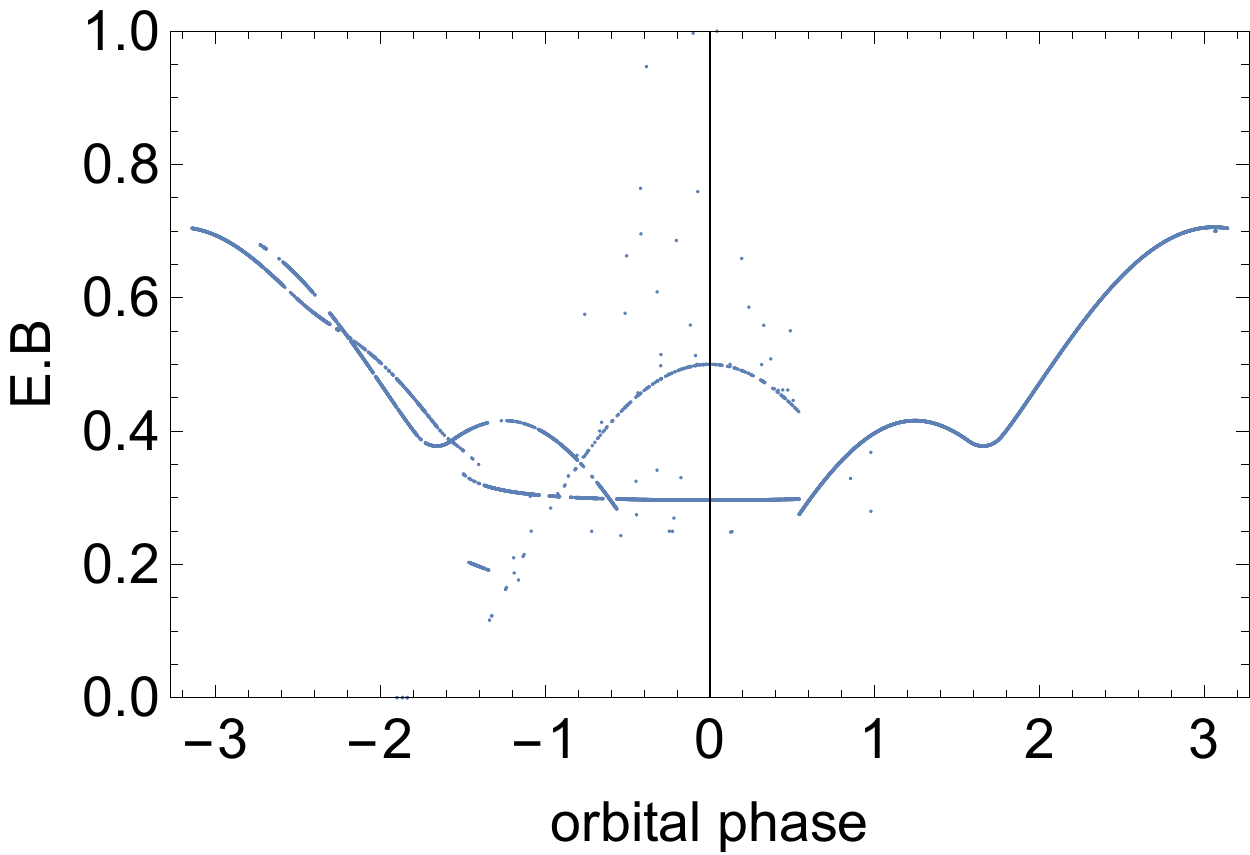} 
\caption{Interaction of the orthogonal dipoles. One dipole is aligned with the  rotational axis, another is in the orbital plane. Left panel: Expected emission pattern (defined as a direction of the \Bf\ at the point where $E_\parallel$ is maximal) as as function of polar angle $\theta$ and azimuthal angle $\phi$. Right panel: maximal value of $E_\parallel$  as a function of orbital phase of the two {\NS}s.}
 \label{emissionofthetaphi}
\end{figure}

 At the present state, we cannot predict radio emission properties given the macroscopic \EM\ structure of the \ms. Hence, we must resolve to probable emission indicators to predict the light curves. Let us choose the value of the parallel \Ef\ as a proxy for the directional properties of radio emission.  For any configuration we know $\E\cdot\B$, Eq (\ref{EdotB}) and  Fig. \ref{Magnetosph}.  Let us assume that emission is produced along the \Bf\ line at the point of maximal $\E\cdot\B$ within the \ms.  As a function of the polar angle $\theta$ and the azimuthal angle $\phi$ the resulting emission will have a complicated patter, Fig \ref{emissionofthetaphi}.

 In both plots in Fig. \ref{emissionofthetaphi}
 several curves arise because the code is looking  for a global 
maximum but  occasionally gets ``trapped'' in  the local  ones. 
  This also indicates that emission may/could change sporadically, appearing
at different field lines.

Finally, we note that in the 2M-DNS configuration  $B^2-E^2 \geq 0$   (it may become zero at some discreet number of point). 
 
 \subsection{2M-DNS - reconnection flares at corotation}
 
 If the stars are rotating, there is another possibility for the production of radiation due to linking of the two \mss\ at the time when the orbital period matches the pulsar's rotational period - at corotation.  \citep[Magnetospheric interaction in close  RS CVn  binaries, \eg ][happens in this regime of near-corotation, whereby two stars can establish magnetic connections.]{1983ASSL..102..629U}

 Equating \NS\ spin (\ref{ONS}) with the orbital frequency at time $(-t)$ before the merger
 \be
 \Omega  \approx \frac{c^{15/8}}{G^{5/8} (-t)^{3/8} M_{{NS}}^{5/8}}
 \ee
 the corotating occurs at 
 time
 \be
(-t_c)\approx  \frac{c {tNS}^{4/3} B_{{NS}}^{8/3} R_{{NS}}^8}{ {I_{NS}}^{4/3}
   ( G M_{{NS}})^{5/3}} \approx  5 \times 10^8 {\rm \, sec}
   \ee
   At the time $t_c$ the stars are separated by 
   \be
   r_c =  \frac{  t_{{NS}}^{1/3} B_{{NS}}^{2/3} ({ G M_{{NS}}})^{1/3}
   R_{{NS}}^2}{c {{I_{NS}}}^{1/3}}
   \approx 5 \times 10^{8}\, t_{NS,9} ^{1/3} {\rm \, cm} 
   \ee

\begin{figure}[h!]
\includegraphics[width=.49\columnwidth]{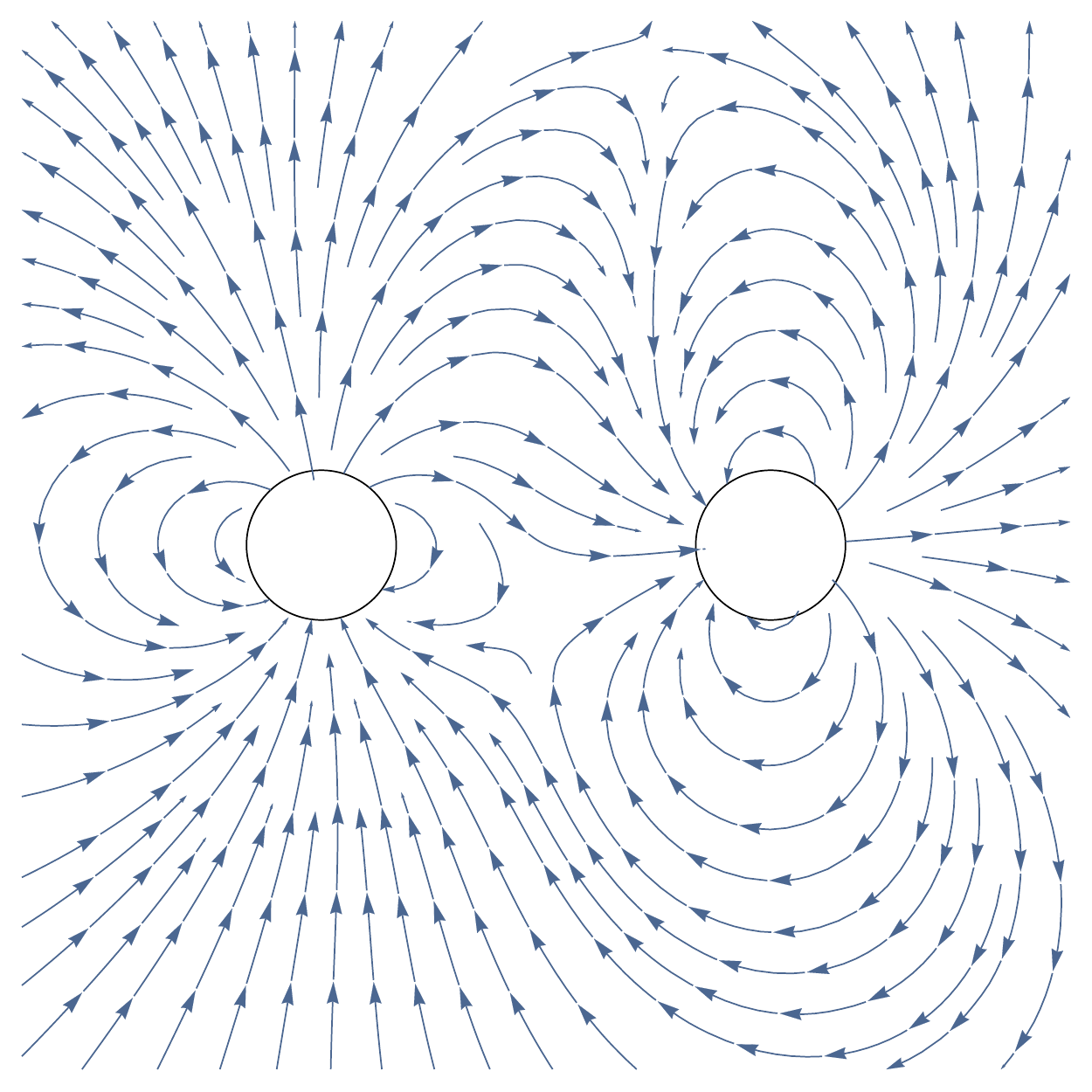} 
\includegraphics[width=.49\columnwidth]{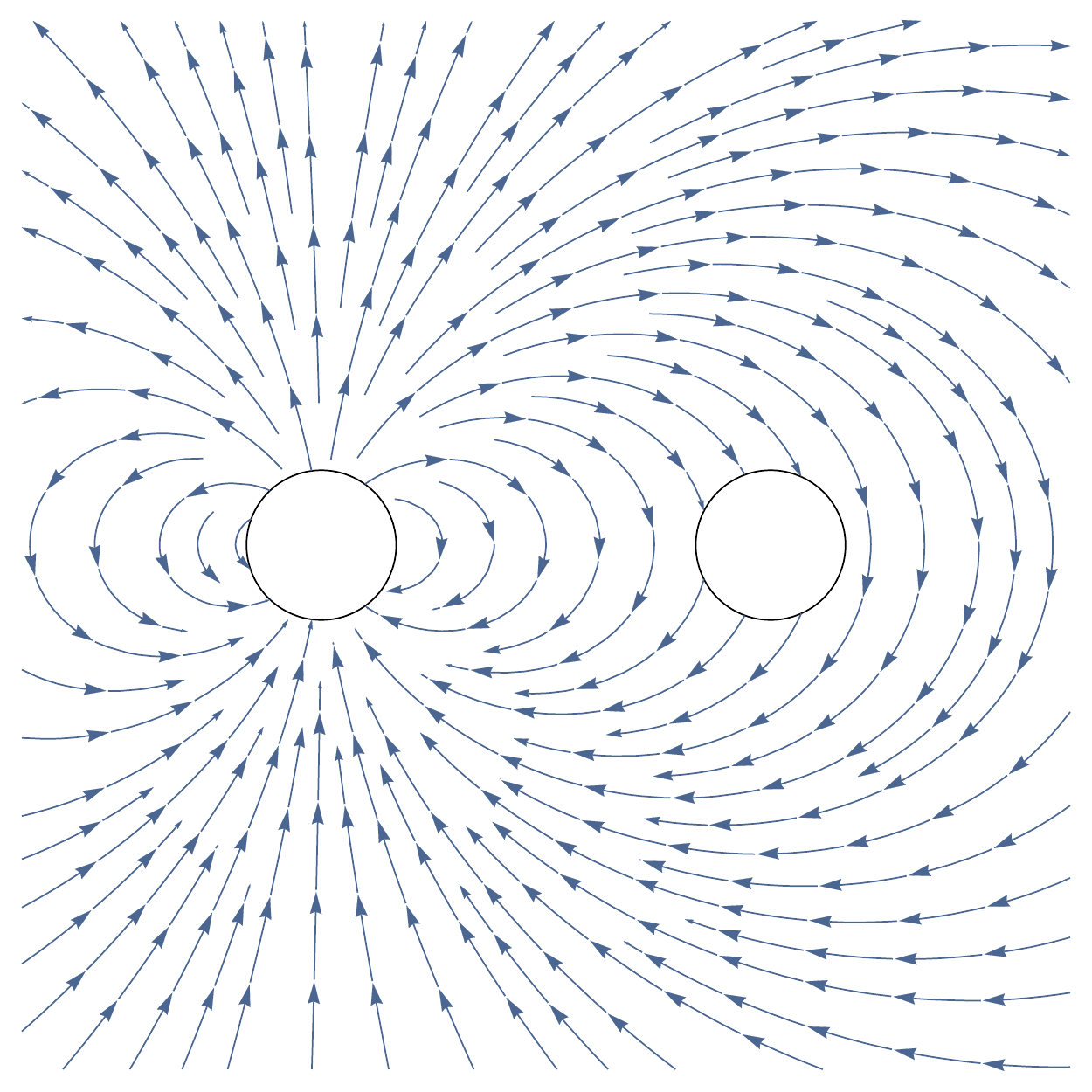}
\caption{Structure of co-rotating  reconnected \mss\  for two orthogonal cases in the $x-z$ plane.}
 \label{BBtot}
\end{figure}

   If two stars are in approximate corotation, there is  time for  magnetospheric field lines to reconnect, creating a link between the two stars.
    We stress that the stars become magnetically connected not due to the diffusive processes in the crust, but due to reconnection in the common \ms, see Fig.  \ref{BBtot}.  

    After two stars established magnetic connection, the common field lines will be  stretched   due to (small) mismatch in the corotation condition.
    This mechanism puts some of the energy of the orbital motion in the \Bf. After a critical value of magnetic bending,  instabilities will disrupt the magnetic bridge between. In the process part of the magnetic energy of the flux tube will be dissipated. 
  
      The total  magnetic energy of a  flux tube connecting two stars can be estimated as
    \be
    E_c \approx \frac{B_{NS}^2 R_{NS}^4}{ 8 r_c} = 
    \frac{c {{I_{NS}}}^{1/3}  B_{{NS}}^{4/3} R_{{NS}}^2}{8 
 ( G M_{NS})^{1/3} t_{{NS}}^{1/3}} =
   3 \times 10^{38} \, t_{NS,9} ^{-1/3}{\rm \, erg}
   \ee
    This is a fairly small amount of energy, unlikely to be detectable (unless a large fraction is put into radio).

 \subsection{A step towards  a plasma model of interacting \mss}
 
 The above considerations were based on the vacuum assumption.  Appearance of parallel {\Ef}s, as well as regions with $E>B$, will lead to pair production. As a result the global plasma dynamics will be (could be)  described by a plasma (MHD or relativistic force-free) approximation. As a step towards plasma models let us estimate  typical current  densities and the corresponding charge density. 
 
 Two lines of reasoning give the bracketing estimates of the plasma density in  the 2M-DNS model.  
 First, the typical current   and  charge  density in the \ms\ can be estimated as
 \ba &&
I \sim c \beta  B_{or} x_0
\nn &&
j \sim \frac{I}{\pi x_0^2} \sim B_{or} \Omega/\pi
\nn &&
n _{GJ}\sim \frac{j}{ e c} = \frac{ B_{or} \Omega}{\pi e c}
\label{nn}
\ea
 which has a clear relation to the Goldreich-Julian density. Density $n_{GJ} $, Eq. (\ref{nn}), is the charge  density we expect  in the \mss\ of interacting \NSs.
 
 Second, in the time-dependent configuration 2M-DNS, 
 equating typical $\partial _t \E$ to the electric current $j$ (in proper units), 
 we find
 \ba && 
 j \propto   B_{NS} \frac{c R_{NS}}{\pi x_0^2} \left( \frac{R_{NS} \Omega}{c}\right)^2 =
 \frac{  B_{or}}{\pi} \frac{G M}{c x_0^2}
  \nn && 
  n_{EM} = \frac{j}{e c} = \frac{ B_{NS}}{\pi ec} \frac{c R_{NS}}{x_0^2} \left( \frac{R_{NS} \Omega}{c}\right)^2=
 \frac{B_{or} \Omega}{\pi ec} \sqrt{\frac{r_G}{x_0}} 
=
 \beta  n _{GJ}
 \label{GJ}
 \ea
where $ r_G =  GM /c^2$,  $\beta = \sqrt{ GM/ (c^2 x_0)}=\sqrt{r_G/x_0}$, $ B_{or} \sim  B_{NS} (R_{NS} /x_0)^{-3}$ is the average \Bf\ in the immediate surrounding of the \NSs, and we used $\Omega \approx \sqrt{G M/x_0^3}$  (recall that $x_0$ is the (semi)-separation between the stars). 
Thus, time-dependence induces a weaker requirement on the charge density, similar to the pulsar \ms\ case.

By the nature of our vacuum approximation, the  expected plasma current is fairly distributed. After the plasma is produces, narrow current sheets may/will be formed, leading  to intermittent, and possibly explosive relativistic  reconnection  \citep{LyutikovUzdensky,2017JPlPh..83f6301L,2017JPlPh..83f6302L},  particle acceleration, and production of the non-thermal radiation. Thus, we expect that magnetospheric interaction produces flares.

\subsection{The last few minutes: possible compactness shroud}
As the  \NSs\ spiral in, the potential and the total luminosity increase. As some point before the merger
 the compactness parameter, 
\be 
l_c = \frac{\sigma_T L_2}{4\pi r^2 m_e c^3}= \frac{2}{3} \frac{e^4 G M_{NS} B_{NS} ^2 R_{NS}^6}{m_e ^3 c^8 r^6}=
\frac{2}{3} \frac{e^4 B_{NS} ^2 R_{NS}^6}{ c^{1/2}(G M_{NS})^{7/2}m_e ^3 c^8 (-t)^{3/2}}
   =3 \times 10^6 (-t)^{-3/2}
   \label{lc}
   \ee
becomes larger than unity. This occurs   at
     \ba &&
     r_c = \frac{e^{2/3} B_{NS}^{1/3} R_{NS} (G M_{NS})^{1/6}}{c^{4/3} m_e^{1/2}}= 5 \times 10^7 {\rm cm}
     \nn &&
   -  t_c= \frac{e^{8/3} B_{NS}^{4/3} R_{NS}^4 }{c^{1/3} m_e^{2} (G M_{NS})^{7/3}}=2 \times 10^4 {\rm sec}.
     \ea
     
    Qualitatively, when the compactness parameter becomes larger than unity a large fraction of energy is converted into pairs that form an optically dense plasma that in turn
may shroud the central \NSs\ \citep{2016MNRAS.461.4435M}.
    
    If the shrouding indeed occurs, this will be a bad news for  a possibility to observe the precursors, since (i) the luminosity of the expected GRB-like outflow from the shroud will be small - of the total luminosity (\ref{L2}), which is already small, most energy will be spent on flow acceleration and not on  the production of emission \citep[recall the low efficiency problem of early theories of GRB outflows][]{Goodman86}; (ii) the shroud will redistribute possibly directed emission from gaps/reconnection cites into isotropic fireball, preventing orbital modulation; (iii) shrouding is likely to pollute the gaps and prevent generation of coherent radio emission.
    
    It is far from clear that shrouding would indeed occur since the estimate (\ref{lc}): (i)  is based on the total Poynting luminosity, the real \EM\ luminosity will be smaller, resulting in smaller $l_c$;    assumes that most of the luminosity initially  comes out in pair-producing photons with energy $\geq 1$ MeV; assumes that radiation processes  are isotropic - anisotropic particle acceleration in the gaps/current sheets will produce anisotropic photon distribution which will reduce pair production efficiency.

\section{Discussion}

In this paper we argue that an effective ``friction'' of the  of \mss\ due to the orbital motion can revive pair production and lead to generation of \EM\ emission. We demonstrate that the
interaction of the \Bfs\ of the merging \NSs\ can create vacuum gaps, somewhat akin to the outer gaps in pulsar \mss\ \citep{1986ApJ...300..500C,Romain96}, and reconnection sheets.
The total available  potentials  in merging \NSs\ (Eqns. (\ref{L1})-(\ref{L2})) are  larger that what is need to produce pairs (sufficiently close to the merger), these parallel {\Ef}s, and regions with $E>B$, will result in pair production, that will screen them, and reduce the problem to the plasma case of $\E\cdot \B\approx 0$. In the process, this will (may) create particle distributions that are unstable to the generation of radio waves.  

The problem of pulsar radio emission - generation of  radio waves by a relativistic unipolar inductor -  is notoriously complicated. It is a general agreement that parallel {\Ef}s are needed.
Particles will be accelerated in the gaps, and under certain conditions can generate a dense  secondary pair plasma.  Plasma instabilities 
\citep[\eg][]{Melrosebook,1999ApJ...512..804L,Melrose00Review} then may lead to the generation of coherent radio emission.

The active stage of the DNS merger,  when LIGO gets an appreciable signal, lasts for $\sim 100$ seconds \citep{2017PhRvL.119p1101A}; hence  there may be enough time for radio observations, \eg\ with LOFAR. LOFAR can see the whole sky at a given time, especially with its
low frequency antennas (LBAs, 10-90MHz); the high frequency antennas
(HBAs, 110-240MHz) can see the $\sim$ 20 degree-wide field. The data rate
is huge, so that phase correlations are mostly done in real time;
there is a possibility to have data stored using the Transient Buffer
Boards, to be linked to transients' notifications. The low frequency
radio waves can also be delayed by $\sim$  tens of seconds due to plasma
dispersion effects, possibly giving extra time for responses to the
notification of a transient.
    
\section*{Acknowledgments}
This work had been supported NSF grant AST-1306672, DoE grant DE-SC0016369 and
NASA grant 80NSSC17K0757.

ML would like to thank  Dimitrios Giannios, Jason  Hessels,  Stephen Reynolds and Joeri van Leeuven for discussions, and organizers and participants  of the Plasma Physics of Neutron Star Mergers workshop.


\bibliographystyle{apj}
\bibliography{/Users/maxim/Home/Research/BibTex}

\end{document}